\documentstyle[12pt]{article}
\newlength{\extraspace}
\newlength{\extraspaces}
\catcode`\@=11
\def\numberbysection{\@addtoreset{equation}{section}
\def\theequation{\arabic{section}.\arabic{equation}}}
\newcommand{\be}{\begin{equation}
\addtolength{\abovedisplayskip}{\extraspaces}
\addtolength{\belowdisplayskip}{\extraspaces}
\addtolength{\abovedisplayshortskip}{\extraspace}
\addtolength{\belowdisplayshortskip}{\extraspace}}
\newcommand{\ee}{\end{equation}}
\newcommand{\ba}{\begin{eqnarray}
\addtolength{\abovedisplayskip}{\extraspaces}
\addtolength{\belowdisplayskip}{\extraspaces}
\addtolength{\abovedisplayshortskip}{\extraspace}
\addtolength{\belowdisplayshortskip}{\extraspace}}
\newcommand{\ea}{\end{eqnarray}}
\newcommand{\newsection}[1]{
\vspace{7mm}
\pagebreak[3]
\addtocounter{section}{1}
\setcounter{equation}{0}
\setcounter{subsection}{0}
\setcounter{footnote}{0}
\begin{center}
{\large {\bf \thesection. #1}}
\end{center}
\nopagebreak
\medskip
\nopagebreak
\hspace{3mm}}
\newcommand{\nonu}{\nonumber \\[.5mm]}
\newcommand{\A}{&\!\!\!}

\newcommand{\e}{\, {\rm e}}

\newcommand{\VEV}[1]{\left\langle {#1} \right\rangle}

\newcommand{\figone}{
\begin{figure}[tb]
\setlength{\unitlength}{0.0125in}
\begin{picture}(280,85)(-55,745)
\thicklines
\put(100,800){\circle{40}}
\put( 80,800){\line(-1, 0){ 20}}
\put( 93,755){\makebox(0,0)[lb]
    {\raisebox{0pt}[0pt][0pt]{\twlrm (a)}}}
\put(260,800){\circle{40}}
\put(240,800){\line(-1, 0){ 20}}
\put(280,800){\line( 1, 0){ 20}}
\put(253,755){\makebox(0,0)[lb]
    {\raisebox{0pt}[0pt][0pt]{\twlrm (b)}}}
\end{picture}
\caption{One-loop divergent diagrams.}
\label{figureone}
\vspace{5mm}
\end{figure}}
\begin{document}
\addtolength{\baselineskip}{.7mm}
\thispagestyle{empty}
\begin{flushright}
TIT/HEP--254 \\
STUPP--94--136 \\
{\tt hep-th/9405072} \\
May, 1994
\end{flushright}
\vspace{2mm}
\begin{center}
{\Large{\bf Dilaton Gravity in $2+\epsilon$ Dimensions}} \\[15mm]
{\sc Shin-ichi Kojima},\footnote{
\tt e-mail: kotori@phys.titech.ac.jp} \hspace{5mm}
{\sc Norisuke Sakai}\footnote{
\tt e-mail: nsakai@phys.titech.ac.jp} \\[3mm]
{\it Department of Physics, Tokyo Institute of Technology \\[2mm]
Oh-okayama, Meguro, Tokyo 152, Japan} \\[4mm]
and \\[4mm]
{\sc Yoshiaki Tanii}\footnote{
\tt e-mail: tanii@th.phy.saitama-u.ac.jp} \\[3mm]
{\it Physics Department,
Saitama University, Urawa, Saitama 338, Japan} \\[15mm]
{\bf Abstract}\\[5mm]
{\parbox{13cm}{\hspace{5mm}
Quantum theory of dilaton gravity is studied in $2+\epsilon$
dimensions. Divergences are computed and renormalized at one-loop
order. The mixing between the Liouville field and the dilaton field
eliminates $1/\epsilon$ singularity in the Liouville-dilaton
propagator. This smooth behavior of the dilaton gravity theory in
the $\epsilon \rightarrow 0$ limit solves the oversubtraction
problem which afflicted the higher orders of
the Einstein gravity in $2+\epsilon$ dimensions. As a nontrivial
fixed point, we find a dilaton gravity action which can be transformed
to a CGHS type action.
}}
\end{center}
\vfill
\newpage
\setcounter{section}{0}
\setcounter{equation}{0}
\numberbysection
%
%
%
\newsection{Introduction}
Ultraviolet divergences and the apparent nonrenormalizability
have been a major source of difficulties to formulate a consistent
quantum theory of gravity. Simple dimensional analysis shows that
gravity is power counting renormalizable in two dimensions.
Therefore it is useful to consider
the quantum theory of gravity near two dimensions
and to make an analytic continuation to higher dimensions
by means of the $\epsilon$ expansion \cite{WEI}--\cite{CD}.
Recently there has been a progress in this
($2+\epsilon$)-dimensional approach for quantum gravity
\cite{KN}--\cite{KST}. The dynamics of the Liouville mode has been
better understood and an ultraviolet fixed point has been found.
\par
Initiated by the work on the black hole evaporation \cite{CGHS},
much efforts have been devoted to study gravity interacting with
a dilaton field and matter fields, especially in two dimensions
\cite{RUTS}--\cite{MASATAUC}. As a toy model imitating the
spherically symmetric situation, two-dimensional models are
considered in most of the works. The semi-classical approximation
has often been used to study the Hawking radiation and the
black hole evaporation in the model \cite{CGHS}--\cite{BDDO},
which is often blamed to be the possible origin of diseases in this
problem. Therefore it is very desirable to have a quantum theory of
the dilaton gravity aiming at the higher dimensional situation.
\par
The purpose of the present work is to study the quantum theory of
the dilaton gravity in $2+\epsilon$ dimensions, which can be used
as a starting point for the higher dimensional quantum theory of
gravity. As a bonus from the study of the dilaton
gravity, we find that the dilaton gravity can solve the
oversubtraction problem \cite{KKN}, which has been a basic obstacle
to renormalization at higher orders in
the usual ($2+\epsilon$)-dimensional approach to quantum gravity.
We obtain divergences and beta functions to one-loop which exhibit
a nontrivial fixed point. The fixed point is
ultraviolet stable for the gravitational coupling constant $G$, if
$\epsilon > 0$ and $N < 24$.
However, it is not ultraviolet stable for the strength of the dilaton
coupling function.
We find that the fixed point theory can be transformed to an action of
the usual CGHS type \cite{CGHS}.
\par
It has been known for some time that there is a subtlety in the
($2+\epsilon$)-dimensional approach to quantum gravity because of
the following intrinsic problem of gravity in two spacetime
dimensions. In the limit of two dimensions, the usual Einstein
action becomes a topological invariant, which is dynamically
meaningless. This peculiarity of the Einstein action at two
dimensions implies that the Liouville field $\rho$ corresponding
to the conformal degree of freedom does not appear in the Einstein
action. However, quantum theory possesses a conformal anomaly,
which resuscitates the Liouville field to play a dynamical role.
One way to see this nontrivial dynamics of the Liouville field is
to consider the quantum gravity in $2+\epsilon$ dimensions.
The Liouville field has a kinetic term of order
$\epsilon$ in $2+\epsilon$ dimensions,
since the Einstein action becomes a topological invariant only
at two dimensions. Therefore the Liouville
field propagator has a $1/\epsilon$ singularity, which is the
origin of a number of subtleties in the ($2+\epsilon$)-dimensional
approach to quantum gravity. One finds divergences
associated to the traceless mode $h_{\mu\nu}$ (graviton), but
no divergences corresponding to the Liouville field kinetic term
at one-loop order. On the other hand, general covariance dictates
that the $1/\epsilon$ counter term for the graviton $h_{\mu\nu}$
inevitably accompanies a finite amount of Liouville field kinetic
term. Since the Liouville field kinetic term is of order $\epsilon$
at the tree level, this finite counter term is an oversubtraction.
\par
At one-loop order, we can perform this oversubtraction as is
dictated by the general coordinate invariance in the $2+\epsilon$
dimensions. However, it has
been noted that the multiple insertion of this finite counter
term produces extra singularities in $1/\epsilon$, because of
the $1/\epsilon$ singularity of the Liouville field propagator.
These extra singularities cannot be removed by any conventional
renormalization procedure.
An unconventional procedure has been proposed to resum
infinitely many diagrams in defining a bare coupling before the
renormalization, but a concrete procedure including higher orders
is still to be worked out \cite{KKN}.
\par
Since this difficulty is associated with an intrinsic problem of the
topological nature of the Einstein action at two dimensions, it
seems difficult to cure the problem. In fact, this difficulty has
even led to a proposal to abandon the general coordinate invariance
of quantum gravity in $2+\epsilon$ dimensions. Namely the action
should be only approximately invariant under general coordinate
transformations at large distances, but is not invariant at short
distances \cite{KKNS}. We admit that this ambitious possibility is
not ruled out by any experimental facts. However, it seems
to us more natural and satisfactory if we can maintain general
coordinate invariance while keeping the idea of expanding the
quantum theory around two dimensions. As we shall see, the dilaton
gravity in $2+\epsilon$ dimensions offers an alternative
possibility to define the quantum theory of Einstein gravity in
higher dimensions by overcoming this oversubtraction problem.
\par
In sect.\ 2, we show that a general model of the dilaton gravity in
$2+\epsilon$ dimensions can be reduced to a standard form
containing an arbitrary function of dilaton coupling to matter
fields. We also explain that
the quantum theory of gravity in higher dimensions can be defined
by means of the dilaton gravity instead of
the usual Einstein gravity in $2+\epsilon$ dimensions.
In sect.\ 3, the standard form of the dilaton gravity in
$2+\epsilon$ dimensions is quantized and our solution to the
oversubtraction problem is pointed out.
In sect.\ 4, one-loop divergences are computed in a general model for
the dilaton gravity.
In sect.\ 5, beta functions are obtained. We find a
 nontrivial fixed point and examine its stability.
Sect.\ 6 is devoted to a discussion.
%
%
\newsection{Dilaton gravity in $2+\epsilon$ dimensions}
The Einstein action for the metric $g_{\mu\nu}$ has been
considered in $d=2+\epsilon$ dimensions together with a number of
free matter fields $X^i$ $(i=1, \cdots, N)$ \cite{WEI}--\cite{KKNS}
\be
S = \int d^d x \sqrt{-g} \biggl[ {\mu^\epsilon \over 16\pi G}
R^{(d)}
- {1 \over 2} g^{\mu\nu} \partial_\mu X^i \partial_\nu X^j
\delta_{ij} \biggr],
\ee
where $G$ and $\mu$ are the renormalized gravitational constant
and the renormalization scale respectively.
String theory and other models have inspired the notion of the
spacetime-dependent gravitational constant which is described by
a field $\phi$ called dilaton.
In order to allow an arbitrary interaction of the dilaton
and to be able to renormalize the theory around two spacetime
dimensions, we shall start from the following general action
\cite{TSE} containing only parameters that become dimensionless
in the limit of two dimensions
\ba
S \A =\A \int d^d x \sqrt{-g} \biggl[ {\mu^\epsilon \over 16\pi G}
R^{(d)} L(\phi, X)
- {1 \over 2} g^{\mu\nu} \partial_\mu \phi \partial_\nu \phi
G_{\phi\phi}(\phi, X) \nonu
\A \A - g^{\mu\nu} \partial_\mu \phi \partial_\nu X^j
G_{\phi j}(\phi, X)
- {1 \over 2} g^{\mu\nu} \partial_\mu X^i \partial_\nu X^j
G_{ij}(\phi, X) \biggr],
\label{generalaction}
\ea
which contains four arbitrary functions $L, G_{\phi\phi}, G_{\phi j},
G_{ij}$ of $\phi$ and $X^i$.
Since we consider matter fields $X^i$ to be free apart from
possible gravitational interactions with metric and dilaton,
we shall demand the action to be
invariant under the $N$-dimensional Euclidean transformations
(translations and rotations) among matter fields $X^i$ (target space).
This requirement distinguishes the matter fields $X^i$ from
the dilaton field $\phi$.
Then the action is restricted to
\be
S =\int d^d x \sqrt{-g} \biggl[ {\mu^\epsilon \over 16\pi G}
R^{(d)} L(\phi)
- {1 \over 2} g^{\mu\nu} \partial_\mu \phi \partial_\nu \phi
\Psi(\phi)
- {1 \over 2} g^{\mu\nu} \partial_\mu X^i \partial_\nu X^j
\delta_{ij} \e^{-2\Phi(\phi)} \biggr].
\label{action}
\ee
\par
Within the class of models of the form (\ref{action}),
we still have a freedom of field redefinitions.
There are three kinds of possible transformations.
The first one is the rigid rescaling of matter fields
\be
X^i \rightarrow a X^i,
\label{xrescale}
\ee
where $a$ is a constant.
The second one is the local Weyl rescaling of the metric
\be
g_{\mu\nu} \rightarrow \e^{-2\Lambda(\phi)} g_{\mu\nu},
\label{localweylrescale}
\ee
where $\Lambda$ is a function of the dilaton $\phi$.
The third one is an arbitrary field redefinition of the dilaton
with a function $f$ of the dilaton $\phi$
\be
\phi \rightarrow f(\phi).
\label{dilatonredef}
\ee
Using the freedom corresponding to the rigid rescaling
(\ref{xrescale}), we can always make $\Phi(0) = 0$.
With this choice, the first freedom is fixed completely.
The local Weyl rescaling (\ref{localweylrescale}) can be used to
fix one of the functions $L, \Psi$, or $ \Phi$.
The function $\Psi$ changes by terms of order
$\epsilon^0$ by the local Weyl rescaling, whereas
$L, \Phi$ change only by terms of order $\epsilon$.
Therefore we shall choose to fix the function $\Psi$ by means of
the local Weyl rescaling.
We can find the local Weyl rescaling $\Lambda$ which transforms a
generic model to the model with $\Psi(\phi)=0$, and is finite as
we let $\epsilon \rightarrow 0$.
Let us note that this choice fixes only $\Lambda'(\phi)$, namely
the nonzero modes of $\Lambda(\phi)$.
Finally the arbitrary redefinition (\ref{dilatonredef})
of the dilaton field $\phi$ can
be used to fix the form of the function $L(\phi)$.
We shall choose $L(\phi)=\exp(-2\phi)$.
This fixes only nonzero modes of $f(\phi)$.
The zero modes of $\Lambda(\phi)$ and $f(\phi)$ are used
to fix coefficients of two reference operators, which will be
introduced into the action later.
We shall discuss reference operators in sect.\ 6.
With these choices of $a, \Lambda(\phi), f(\phi)$,
we can fix $L(\phi), \Psi(\phi)$ in the action (\ref{action})
and obtain the following standard form of dilaton gravity
\be
S = \int d^d x \sqrt{-g} \biggl[ {\mu^\epsilon \over 16\pi G}
R^{(d)} \e^{-2\phi}
- {1 \over 2} g^{\mu\nu} \partial_\mu X^i \partial_\nu X^j
\delta_{ij} \e^{-2\Phi(\phi)} \biggr],
\label{standardform}
\ee
where $\Phi(0) = 0$.

Let us note that the field redefinition generally produces Jacobian
factors in the path integral measure in the usual regularization
schemes.
 In $2+\epsilon$ dimensions, however, these Jacobian factors do not
contribute to the effective action since they are proportional to
$\delta^d(0) = 0$ which are discarded in the analytic
regularization such as the ($2+\epsilon$)-dimensional approach
\cite{BRZJLG}.
\par
It is important to realize that the dilaton gravity is equivalent
to the Einstein gravity with one extra scalar field.
In order to demonstrate this point, let us take
the dilaton gravity without matter
fields in $2+\epsilon$ dimensions
\be
S ={\mu^\epsilon \over 16\pi G}
\int d^d x \sqrt{-g} R^{(d)} \e^{-2\phi}.
\label{dilwithoutmatter}
\ee
If we make a field redefinition involving a local Weyl
rescaling $g_{\mu\nu} \rightarrow \e^{4\phi / \epsilon} g_{\mu\nu}$,
we find
\be
S = {\mu^\epsilon \over 16\pi G} \int d^d x \sqrt{-g} \biggl[ R^{(d)}
- {4(1+\epsilon) \over \epsilon}
g^{\mu\nu} \partial_\mu \phi \partial_\nu \phi \biggr].
\label{redaction}
\ee
This clearly shows that the dilaton gravity is equivalent to the
Einstein gravity with one scalar field. However, the coefficient
of the $\phi$ kinetic term is singular for $\epsilon \rightarrow 0$
in the action (\ref{redaction}).
If we are precisely at two dimensions, we cannot
transform the Einstein action, which is a topological invariant
given by the first term of eq.\ (\ref{redaction}),
back into the dilaton gravity action (\ref{dilwithoutmatter}).
Therefore the dilaton gravity is different from the Einstein gravity
precisely at two dimensions, even though they are equivalent in
dimensions other than two.
Since we are going to use an expansion in $\epsilon = d-2$,
we should only consider field redefinitions which keep
the action regular for $d \rightarrow 2$.
The dilaton gravity and the Einstein gravity in $2+\epsilon$
dimensions give different divergences around two dimensions.
Therefore the Einstein gravity
and the dilaton gravity in ($2+\epsilon$)-dimensional approach
give inequivalent quantum gravity theories in higher dimensions,
even though they are classically equivalent in
dimensions other than two.
\par
Since the dilaton gravity is equivalent to the Einstein gravity
in higher dimensions,
the dilaton gravity in $2+\epsilon$ dimensions is as
legitimate as the Einstein gravity in $2+\epsilon$ dimensions,
in order to define the quantum gravity in higher dimensions.
The dynamical content of the
Einstein action does not have a smooth limit as
$\epsilon \rightarrow 0$.
This is precisely the origin of a number of problems in the
usual ($2+\epsilon$)-dimensional approach, such as the oversubtraction
problem \cite{KKN}.
On the other hand, the dilaton gravity action still has a dynamical
content even at two dimensions and has a smooth limit as
$\epsilon \rightarrow 0$.
This smooth behavior of dilaton gravity in $\epsilon \rightarrow 0$
suggests that the dilaton gravity is more natural than
the Einstein gravity to define the quantum gravity theories in
higher dimensions in the ($2+\epsilon$)-dimensional approach.
\par
It is customary to incorporate matter fields as free fields.
One should note that the notion of free fields is invariant under
the local Weyl rescaling of the form (\ref{localweylrescale})
only at two dimensions. Since we need
to consider the Weyl rescaling in order to relate the dilaton
gravity action and the Einstein action, we are
led to consider the free matter fields $X^i$ to interact
with dilaton in addition to the metric as in eq.\ (\ref{standardform}).
This is quite natural, since the dilaton is better regarded as a
part of gravity.
\par
A number of different forms of dilaton gravity has been proposed
so far. A popular form of the action for the dilaton gravity is the
CGHS type action \cite{CGHS}. It can be obtained from our standard
action (\ref{standardform}) by a local Weyl rescaling
$g_{\mu\nu} = \e^{-2\phi}g'_{\mu\nu}$.
The transformed action becomes in $2+\epsilon$ dimensions
\ba
S_{\rm CGHS} \A = \A
\int d^d x \sqrt{-g'} \biggl[ {\mu^\epsilon \over 16\pi G}
\biggl( R'^{(d)}
+ (\epsilon+1)(\epsilon+4)g'^{\mu\nu} \partial_\mu \phi
\partial_\nu \phi\biggr) \e^{-(2+\epsilon)\phi} \nonu
\A \A - {1 \over 2} g'^{\mu\nu} \partial_\mu X^i
\partial_\mu X^i \delta_{ij} \e^{-2\Phi(\phi)-\epsilon \phi} \biggr].
\label{cghsaction}
\ea
%
%
\newsection{Gauge fixing and quantization}
We shall use the background field method \cite{DEWITT}, \cite{TV}
to compute one-loop divergences.
We define an expansion parameter $\kappa$ as
\be
\kappa^2 = {16\pi G \over \mu^\epsilon}.
\label{kappa}
\ee
Let us decompose the metric $g_{\mu\nu}$ into the traceless
field $h_{\mu\nu}$ and the Liouville field $\rho$
and introduce the background metric $\hat g_{\mu\nu}$
\be
g_{\mu\nu} = \tilde g_{\mu\nu} \e^{-2\rho}
=\hat g_{\mu\lambda} (\e^{\kappa h})^\lambda{}_\nu
\e^{-2\rho}.
\label{metricdec}
\ee
Our standard action (\ref{standardform}) for the dilaton gravity
becomes
\ba
S
\A = \A {\mu^\epsilon \over 16\pi G} \int d^d x \sqrt{-\tilde g}
\e^{-\epsilon\rho} \biggl[ \tilde R^{(d)} \e^{-2\phi}
+ \epsilon(\epsilon+1) \tilde g^{\mu\nu} \partial_\mu \rho
\partial_\nu \rho \e^{-2\phi} \nonu
\A \A + 4 (\epsilon+1) \tilde g^{\mu\nu} \partial_\mu \rho
\partial_\nu \phi \e^{-2\phi}
- {1 \over 2} \tilde g^{\mu\nu} \partial_\mu X^i
\partial_\nu X^i \delta_{ij} \e^{-2\Phi(\phi)} \biggr],
\label{dilatonaction}
\ea
where $i, j = 1, \cdots, N$.
We also introduce the background fields
of the Liouville field, the dilaton field and the matter fields
which are denoted by putting a hat $\, \hat{} \, $
\be
\rho=\hat \rho+\kappa \rho_q, \quad
\phi = \hat \phi + \kappa \phi_q, \quad
X^i = \hat X^i + X^i_q.
\ee
In the following we will omit the suffix $q$ of the quantum fields
$\rho_q, \phi_q$, $X^i_q$ for simplicity.
We use the background vielbein $\hat e_\mu{}^\alpha$
to convert world indices $\mu, \nu, \cdots$ to local Lorentz indices
$\alpha, \beta, \cdots$.
\par
To fix the gauge we introduce the Faddeev-Popov ghosts
$b_\alpha, c^\alpha$ and the Nakanishi-Lautrup auxiliary field
$B_\alpha$. The BRST transformations are given by
\ba
\delta_{\rm B} h_{\alpha\beta} \A = \A
\hat D_\alpha c_\beta + \hat D_\beta c_\alpha
-{2 \over d}\eta_{\alpha\beta} \hat D_\gamma c^\gamma + \cdots, \quad
\delta_{\rm B} \rho = - {1 \over d} \hat D_\alpha c^\alpha
+ \cdots, \nonu
\delta_{\rm B} \phi \A = \A c^\alpha \partial_\alpha \hat\phi
+ \cdots, \quad
\delta_{\rm B} b_\alpha = i B_\alpha, \quad
\delta_{\rm B} B_\alpha = 0,
\label{brst}
\ea
where the dots represent terms quadratic and of higher powers in
the quantum fields. The gauge fixing term and the ghost action are
given by
$S_{\rm GF+FP} = \int d^d x \delta_{\rm B} (-ib^\alpha F_\alpha)$
for a gauge function $F_{\alpha}$ \cite{KU}.
We use the following gauge function to eliminate
the mixing between $h_{\mu\nu}$ and other fields
\be
F_\alpha
= \sqrt{-\hat g} \e^{-2\hat\phi} \left( \hat D^\beta h_{\beta\alpha}
  + \hat D_\alpha ( \epsilon \rho + 2 \phi )
  + {1 \over 2} B_\alpha \right).
\label{dilatongaugefunc}
\ee
The total action quadratic in the quantum fields is given by
\ba
S_{\rm tot}^{(2)}
\A = \A S^{(2)} + S^{(2)}_{\rm GF+FP} \nonu
\A = \A \int  d^d x \sqrt{-\hat g} \e^{-2\hat\phi} \biggl[
{1 \over 2} B'^\alpha B'_\alpha
-{1 \over 4} \hat D_\mu h_{\alpha\beta} \hat D^\mu h^{\alpha\beta}
+ {1 \over 2} \epsilon (\epsilon+2) \hat g^{\mu\nu} \partial_\mu \rho
  \partial_\nu \rho \nonu
\A \A + 2 (\epsilon+2) \hat g^{\mu\nu} \partial_\mu \rho
  \partial_\nu \phi
- 2 \hat g^{\mu\nu} \partial_\mu \phi \partial_\nu \phi
- {1 \over 2} \hat R^{(d)}_{\mu\rho\sigma\nu} h^{\mu\nu} h^{\rho\sigma}
+ {1 \over 2} (\epsilon \rho + 2 \phi)^2 \hat R^{(d)} \nonu
\A \A + (\epsilon \rho + 2 \phi) h^{\mu\nu} \hat R^{(d)}_{\mu\nu}
- 2 h^{\mu\lambda} \hat D_\nu h^\nu{}_\lambda \partial_\mu \hat\phi
- 2 (\epsilon \rho + 2 \phi) \hat D_\nu h^{\mu\nu}
  \partial_\mu \hat\phi \nonu
\A \A - 4 (\epsilon+1) (\epsilon \rho + 2 \phi) \hat g^{\mu\nu}
  \partial_\mu \rho \partial_\nu \hat\phi
- 4 (\epsilon+1) h^{\mu\nu} \partial_\mu \rho
  \partial_\nu \hat\phi + \mbox{ ghost terms} \biggr] \nonu
\A \A + S_{\rm matter}^{(2)},
\ea
where $B'_\alpha = B_\alpha + \hat D^\beta h_{\beta\alpha}
+ \partial_\alpha ( \epsilon \rho + 2 \phi)$
is a shifted auxiliary field. One can easily see
that the terms quadratic in the Liouville field $\rho$ are of order
$\epsilon$ similarly to the case of the Einstein action.
However, it is important to note that the mixing of the Liouville
field $\rho$ with the dilaton field $\phi$ is of order $\epsilon^0$.
Therefore the kinetic term of the Liouville field has to be considered
together with the dilaton field.
\par
By inverting the kinetic term of various fields, we obtain
propagators. The nonvanishing propagators are given by
\ba
\A\A \VEV{h_{\alpha\beta}(x)h_{\gamma\delta}(y)}
= \left( \eta_{\alpha\gamma} \eta_{\beta\delta}
+ \eta_{\alpha\delta} \eta_{\beta\gamma} - {2 \over d}
\eta_{\alpha\beta} \eta_{\gamma\delta} \right) \Delta_F(x-y), \nonu
\A\A \VEV{\rho(x)\rho(y)}
= - {1 \over 2(\epsilon+1)(\epsilon+2)} \, \Delta_F(x-y), \qquad\!\!\!
\VEV{\rho(x)\phi(y)} =  - {1 \over 4(\epsilon+1)} \, \Delta_F(x-y), \nonu
\A\A \VEV{\phi(x)\phi(y)}
= {\epsilon \over 8(\epsilon+1)} \, \Delta_F(x-y), \qquad
\VEV{c^\alpha(x)b_\beta(y)}
 =  - i \delta_\beta^\alpha \, \Delta_F(x-y), \nonu
\A\A \VEV{X^i(x)X^j(y)} = \delta^{ij} \, \Delta_F(x-y), \qquad
\Delta_F(x-y)  =  - i \int {d^d p \over (2\pi)^d} {1 \over p^2}
\e^{i p \cdot (x-y)}.
\label{propagator}
\ea
All the propagators including that for the Liouville field have
well-defined limits as $\epsilon \rightarrow 0$.
This nonsingular behavior is a direct consequence of the mixing
between the Liouville field $\rho$ and the dilaton field $\phi$.
Although the Liouville field kinetic term alone vanishes in the
$\epsilon \rightarrow 0$ limit, the dilaton field mixes with
the Liouville field nontrivially even in the two-dimensional limit.
Therefore the Liouville field $\rho$ and the dilaton field $\phi$
should be considered as inseparable parts of a single entity
in the dilaton gravity.
\par
In the case of the Einstein action in the ($2+\epsilon$)-dimensional
approach, the propagator of the Liouville field is singular.
Since there are divergences proportional to the scalar curvature,
the general coordinate invariance with respect to the physical
background metric
$\bar g_{\mu\nu}=\hat g_{\mu\nu} \e^{-2\hat \rho}$
dictates that the counter term
contains the Liouville field kinetic term of order $\epsilon^0$
\ba
S_{\rm counter} \A \propto \A
\int d^dx {1 \over \epsilon} \sqrt{-\bar g} \bar R^{(d)} \nonu
\A = \A \int d^dx \sqrt{-\hat g} \e^{-\epsilon \rho}
\biggl[{1 \over \epsilon}\hat R^{(d)}
+(1+\epsilon) \hat g^{\mu\nu}
\partial_\mu \rho \partial_\nu \rho \biggr].
\ea
This oversubtraction is the origin of the nontrivial dynamics of the
Liouville field even in the limit of two dimensions. At the same
time, this oversubtraction causes the following problem at higher
orders. If one inserts this counter term into a Liouville field
propagator of any diagram, one finds extra $1/\epsilon$
singularities for each insertion because of the singular behavior
of the Liouville field propagator. The multiple insertions of the
counter term provide more and more singular diagrams. Moreover,
these diagrams are nonlocal and cannot be renormalized in any
conventional way. This is the oversubtraction problem \cite{KKN}.
\par
If we consider the dilaton gravity instead of the Einstein gravity
in the ($2+\epsilon$)-dimensional approach, we find that the
Liouville field should be considered with the dilaton field and
two by two matrix of their propagators (\ref{propagator}) are
nonsingular.
Even though we still need to subtract the finite counter term for
the Liouville kinetic term, we no longer obtain additional divergences
from the multiple insertion of the finite counter terms.
Therefore the oversubtraction problem that has afflicted the
Einstein gravity in $2+\epsilon$ dimensions can be overcome in the
dilaton gravity. Since the dilaton gravity has a smooth limit at
two dimensions unlike the Einstein gravity, the dilaton gravity is
similar to all the other field theoretical models in the $\epsilon$
expansion approach. Therefore we expect that the dilaton gravity
does not have any subtleties in higher orders of the
($2+\epsilon$)-dimensional approach contrary
to the Einstein gravity.
%
%
\newsection{One-loop divergences in a general model}
We shall compute divergences at one-loop order for a general
action instead of the standard one (\ref{standardform})
in order to allow a more flexible treatment.
Combining all the scalar fields and the Liouville field into
$Y^I=(\rho, \phi, X^i)$, where the index runs $I=(\rho, \phi, i)$,
we obtain a kind of nonlinear sigma model \cite{TSE}
using $\tilde g_{\mu\nu}
=\hat g_{\mu\lambda} (\e^{\kappa h})^\lambda{}_\nu$
\be
S = {\mu^\epsilon \over 16\pi G}
\int d^d x \sqrt{-\tilde g} \left[ \tilde R^{(d)} L(Y)
- {1 \over 2} \tilde g^{\mu\nu} \partial_\mu Y^I
\partial_\nu Y^J G_{IJ}(Y) \right].
\label{actiony}
\ee
Although we do not require the invariance under the Weyl rescaling
of the physical metric $\delta g_{\mu\nu}(x) = -2 \sigma(x)
g_{\mu\nu}(x)$, we demand the invariance under the Weyl rescaling
of the background metric $\hat g_{\mu\nu}$ accompanied by a shift
of $Y^I$
\be
\delta \hat g_{\mu\nu} = -2 \sigma \hat g_{\mu\nu}, \qquad
\delta Y^I = - 2 (\epsilon+1) \sigma G^{IJ} \partial_J L.
\label{bweyl}
\ee
In the case of our standard model for the dilaton gravity,
this transformation of $Y^I$
reduces to $\delta \rho = - \sigma$, $\delta \phi, \delta X^i = 0$.
The invariance under the transformation (\ref{bweyl}) implies
that the action depends on $\hat g_{\mu\nu}$ and $\rho$ only
through a combination $g_{\mu\nu}$ in eq.\ (\ref{metricdec}),
which assures the general coordinate invariance with respect
to the physical metric $g_{\mu\nu}$.
This requirement constrains the function $L$ and the target space
metric $G_{IJ}$ as
\ba
- \epsilon G_{IJ}- 4 (\epsilon+1) D_I \partial_J L = 0, \nonu
- \epsilon L- 2 (\epsilon+1) G^{IJ} \partial_I L \partial_J L = 0.
\label{bwcond}
\ea
We will discuss the general solution of these conditions in the
appendix.
We can define the following general coordinate transformation
\ba
\delta_{\rm G} \tilde g_{\mu\nu} \A = \A
\tilde D_\mu \tilde v_\nu + \tilde D_\nu \tilde v_\mu
- {2 \over \epsilon+2} \, \tilde g_{\mu\nu}
\tilde D_\lambda v^\lambda, \nonu
\delta_{\rm G} Y^I \A = \A v^\mu \partial_\mu Y^I
- {2(\epsilon+1) \over \epsilon+2} \, G^{IJ} \partial_J L
\tilde D_\mu v^\mu,
\label{gct}
\ea
where $\tilde v_\mu = \tilde g_{\mu\nu} v^\nu$.
The action (\ref{actiony}) is invariant under eq.\ (\ref{gct})
when the functions $L$ and $G_{IJ}$ satisfy eq.\ (\ref{bwcond}).
\par
We decompose the fields into background fields and quantum
fields as
\be
\tilde g_{\mu\nu}
= \hat g_{\mu\lambda} (\e^{\kappa h})^\lambda{}_\nu, \qquad
Y^I = \hat Y^I + \kappa \xi^I + O(\kappa^2 \xi^2),
\ee
where $\kappa$ is defined in eq.\ (\ref{kappa}) and $\xi^I$ are
normal coordinates with respect to the target space metric
$G_{IJ}$ \cite{AFM}. Terms quadratic in the quantum fields in
the action (\ref{actiony}) are given by
\ba
S^{(2)}
\A = \A \int  d^d x \sqrt{-\hat g} \biggl[ \hat L \biggl\{
-{1 \over 4} \hat D_\mu h_{\rho\sigma} \hat D^\mu h^{\rho\sigma}
+ {1 \over 2} \hat D_\mu h^{\mu\lambda} \hat D_\nu h^\nu{}_\lambda
- {1 \over 2} \hat R^{(d)}_{\mu\rho\sigma\nu} h^{\mu\nu}
h^{\rho\sigma} \nonu
\A \A - \hat D_\mu ( h^{\mu\lambda} \hat D_\nu h^\nu{}_\lambda )
\biggr\} - {\epsilon \over 8(\epsilon+1)} \xi^I \xi^J
\hat G_{IJ} \hat R^{(d)} + \left( \hat D_\mu \hat D_\nu h^{\mu\nu}
- \hat R^{(d)}_{\mu\nu}
h^{\mu\nu} \right) \xi^I \partial_I \hat L \nonu
\A \A - {1 \over 4} h^{\mu\lambda} h^\nu{}_\lambda
\partial_\mu \hat Y^I \partial_\nu \hat Y^J \hat G_{IJ}
+ h^{\mu\nu} \hat D_\mu \xi^I \partial_\nu \hat Y^J \hat G_{IJ}
- {1 \over 2} \hat D_\mu \xi^I \hat D^\mu \xi^J \hat G_{IJ} \nonu
\A \A + {1 \over 2} \partial_\mu \hat Y^I \partial^\mu \hat Y^J
\hat R_{IKJL} \xi^K \xi^L \biggr],
\ea
where $\hat G_{IJ} = G_{IJ}(\hat Y)$, $\hat L = L(\hat Y)$
and we have used the first condition in eq.\ (\ref{bwcond}).
\par
To fix the gauge we introduce the Faddeev-Popov ghosts
$b_\alpha, c^\alpha$ and the Nakanishi-Lautrup auxiliary field
$B_\alpha$. The BRST transformations are
\ba
\delta_{\rm B} h_{\alpha\beta} \A = \A
\hat D_\alpha c_\beta + \hat D_\beta c_\alpha
-{2 \over \epsilon+2}\eta_{\alpha\beta} \hat D_\gamma c^\gamma
+ \cdots, \nonu
\delta_{\rm B} \xi^I \A = \A
c^\alpha \partial_\alpha \hat Y^I
-{2(\epsilon+1) \over \epsilon+2} \hat D_\alpha c^\alpha
\hat G^{IJ} \partial_J \hat L + \cdots, \nonu
\delta_{\rm B} b_\alpha \A = \A i B_\alpha, \qquad
\delta_{\rm B} B_\alpha = 0,
\ea
where the dots represent terms quadratic and of higher powers in
the quantum fields. To eliminate the mixing of $h_{\mu\nu}$ and
other fields, we use the following gauge function
\be
F_\alpha
= \sqrt{-\hat g} \hat L \left[ \hat D^\beta h_{\beta\alpha}
- {1 \over \kappa} \partial_\alpha \left( {L(Y) \over \hat L}
\right) + {1 \over 2} B_\alpha \right].
\label{func}
\ee
Then the gauge fixing term and the ghost action are given by
\ba
S_{\rm GF+FP}
\A = \A \int d^d x \, \delta_{\rm B} \bigl( - i
b^\alpha F_\alpha \bigr) \nonu
\A = \A \int d^d x \, \sqrt{-\hat g} \hat L \biggl[ \,
{1 \over 2} B'^\alpha B'_\alpha
- {1 \over 2} \left( \hat D^\beta h_{\alpha\beta}
- \partial_\alpha \left( \xi^I \partial_I \ln \hat L
\right) \right)^2 \nonu
\A \A + i b^\alpha \hat D^\beta \hat D_\beta c_\alpha
+ i \hat R^{(d)}_{\alpha\beta} b^\alpha c^\beta
- i b^\alpha \hat D_\alpha \left( c^\beta \partial_\beta
\ln \hat L \right) + \cdots \, \biggr],
\label{gffp}
\ea
where $B'_\alpha$ is a shifted auxiliary field and we have used
the second condition in eq.\ (\ref{bwcond}). The total action is
\ba
S_{\rm tot}^{(2)}
\A = \A S^{(2)} + S^{(2)}_{\rm GF+FP} \nonu
\A = \A \int  d^d x \sqrt{-\hat g} \biggl[ \hat L \biggl\{
-{1 \over 4} \hat D_\mu h_{\rho\sigma} \hat D^\mu h^{\rho\sigma}
- {1 \over 2} \hat R^{(d)}_{\mu\rho\sigma\nu}
h^{\mu\nu} h^{\rho\sigma} \nonu
\A \A - \hat D_\mu \left( h^{\mu\lambda}
\hat D_\nu h^\nu{}_\lambda \right)
- \hat R^{(d)}_{\mu\nu} h^{\mu\nu} \xi^I \partial_I \ln \hat L
- \hat D_\mu h^{\mu\nu} \xi^I \partial_I \ln \hat L
\partial_\nu \ln \hat L \nonu
\A \A - {1 \over 2} \hat g^{\mu\nu}
\partial_\mu \left( \xi^I \partial_I \ln \hat L \right)
\partial_\nu \left( \xi^J \partial_J \ln \hat L \right)
+ {1 \over 2} B'^\alpha B'_\alpha
+ \mbox{ ghost terms} \biggr\} \nonu
\A \A - {\epsilon \over 8(\epsilon+1)} \xi^I \xi^J \hat G_{IJ}
\hat R^{(d)}
- {1 \over 4} h^{\mu\lambda} h^\nu{}_\lambda
\partial_\mu \hat Y^I \partial_\nu \hat Y^J \hat G_{IJ}
+ h^{\mu\nu} \hat D_\mu \xi^I \partial_\nu \hat Y^J
\hat G_{IJ} \nonu
\A \A - {1 \over 2} \hat D_\mu \xi^I \hat D^\mu \xi^J \hat G_{IJ}
+ {1 \over 2} \partial_\mu \hat Y^I \partial^\mu \hat Y^J
\hat R_{IKJL} \xi^K \xi^L \biggr].
\label{totalaction}
\ea
\par
It is better to remove interaction terms with two derivatives.
We rescale
\be
h_{\alpha\beta} \rightarrow
\hat L^{-{1 \over 2}} h_{\alpha\beta}, \qquad
B'_\alpha \rightarrow \hat L^{-{1 \over 2}} B'_\alpha, \qquad
b_\alpha \rightarrow \hat L^{-1} b_\alpha.
\label{rescale}
\ee
To reassemble the terms quadratic in $\xi^A$, we define the
effective target space metric $G'_{IJ}$ as
\be
G'_{IJ} =  G_{IJ} + {\partial_I L \partial_J L \over L}, \qquad
G'^{IJ} = G^{IJ} - {2(\epsilon+1) \over \epsilon+2}
{\partial^I L \partial^J L \over L},
\ee
where $\partial^I L = G^{IJ} \partial_J L$ and the second condition
in eq.\ (\ref{bwcond}) is used to obtain the inverse $G'^{IJ}$.
The vielbein can be chosen as
\ba
G_{IJ} \A = \A E_I{}^A E_J{}^B \eta_{AB}, \qquad
G'_{IJ} = E'_I{}^A E'_J{}^B \eta_{AB}, \nonu
E'_I{}^A \A = \A
E_I{}^A + a \, {\partial_I L \partial^A L \over L}, \qquad
E'_A{}^I =
E_A{}^I + b \, {\partial_A L \partial^I L \over L},
\label{vielbein}
\ea
where $\partial_A L = E_A{}^I \partial_I L$ and
\ba
a \A = \A {2(\epsilon+1) - \sqrt{2(\epsilon+1)(\epsilon+2)}
\over \epsilon}
= {1 \over 2} + O(\epsilon), \nonu
b \A = \A - {2(\epsilon+1)a \over 2(\epsilon+1)-\epsilon a}
= - {1 \over 2} + O(\epsilon).
\label{abvalues}
\ea
We define the quantum fields with the index $A$ and the
covariant derivative on them as
\be
\xi^A = \xi^I E'_I{}^A,
\label{newxi}
\ee
\ba
\hat D_\mu \xi^A
\A = \A \partial_\mu \xi^A + \partial_\mu \hat Y^I
\hat\Omega_I{}^A{}_B \xi^B, \nonu
\Omega_I{}^A{}_B \A = \A E'_B{}^J \Gamma_{IJ}^{\; K}  E'_K{}^A
+ \partial_I E'_B{}^J E'_J{}^A
- \partial_I L \partial^A L \partial_B L L^{-2}.
\ea
Here $\Gamma_{IJ}^{\; K}$ is the Christoffel connection of the
metric $G_{IJ}$ (not $G'_{IJ}$). Using eqs.\ (\ref{vielbein}),
(\ref{abvalues}) and (\ref{bwcond}) it can be shown that the
the effective spin connection $\Omega_{IAB}$ is not antisymmetric
in $A, B$
\be
\Omega_{IAB} + \Omega_{IBA} = - \partial_I L \partial_A L
\partial_B L L^{-2} + O(\epsilon).
\label{omegasym}
\ee
\par
In terms of the new fields (\ref{rescale}), (\ref{newxi})
the total action (\ref{totalaction}) becomes
\ba
S_{\rm tot}^{(2)}
\A = \A \int  d^d x \sqrt{-\hat g} \biggl[
-{1 \over 4} \hat D_\mu h_{\alpha\beta} \hat D^\mu h^{\alpha\beta}
- {1 \over 2} \hat R^{(d)}_{\alpha\gamma\delta\beta}
h^{\alpha\beta} h^{\gamma\delta} \nonu
\A \A - {1 \over 16} h_{\alpha\beta} h^{\alpha\beta} \left(
2 \partial^\mu \partial_\mu \ln \hat L
+ \partial^\mu \ln\hat L
\partial_\mu \ln\hat L \right) \nonu
\A \A - {1 \over 4} h^{\alpha\gamma} h^\beta{}_\gamma \left(
2 \partial_\alpha \ln\hat L \partial_\beta \ln\hat L
+ \partial_\alpha \hat Y^I \partial_\beta \hat Y^J
\hat G_{IJ} \hat L^{-1} \right)
+ \partial_\alpha h^{\alpha\beta} h_\beta{}^\gamma
\partial_\gamma \ln \hat L \nonu
\A \A - i \hat D^\beta b^\alpha \hat D_\beta c_\alpha
+ i \hat R^{(d)}_{\alpha\beta} b^\alpha c^\beta
+ i \hat D_\alpha b^\alpha c^\beta \partial_\beta \ln \hat L
+ {1 \over 2} B'^\alpha B'_\alpha \nonu
\A \A - {1 \over 2} \hat D_\mu \xi^A \hat D^\mu \xi^B \eta_{AB}
+ {1 \over 2} \xi^A \xi^B  \partial_\mu \hat Y^I
\partial^\mu \hat Y^J \left( \hat R_{IAJB}
- \partial_I \hat L \partial_J \hat L \partial_A \hat L
\partial_B \hat L \hat L^{-3} \right) \nonu
\A \A + h^{\alpha\beta} \partial_\alpha \xi^A \hat E'_{IA}
\partial_\beta \hat Y^I \hat L^{-{1 \over 2}}
+ \cdots \; \biggr],
\label{redtotalaction}
\ea
where we have used eq.\ (\ref{bwcond}) and the dots represent
terms which do not contribute to one-loop divergences.
They are either terms of order $\epsilon$ or terms proportional
to $\xi^A h_{\mu\nu}$ (without derivatives). Non-vanishing
propagators are the same as those in eq.\ (\ref{propagator}),
except those for $\rho, \phi, X^i$ should be replaced by
those for the fields $\xi^A$
\be
\VEV{\xi^A(x)\xi^B(y)}  = \eta^{AB} \, \Delta_F(x-y).
\ee
\par
Divergences of the effective action at one-loop order should
take the form
\be
\Gamma_{\rm div} = \int d^d x \sqrt{-\hat g} \biggl[
\hat R^{(d)} A(\hat Y) + \hat g^{\mu\nu} \partial_\mu
\hat Y^I \partial_\nu \hat Y^J B_{IJ}(\hat Y) \biggr].
\label{divergence}
\ee
To obtain the coefficient $A(\hat Y)$ in the first term we only
need to consider the case $\hat Y^I =$ constant.
Then the action becomes almost the same as the action of the
Einstein gravity in refs.\ \cite{KKN}, \cite{KST} except for the
terms which do not contribute to the divergence.
Therefore, the coefficient of the first term is given by
\be
A = {24-N \over 24\pi\epsilon}.
\label{adiv}
\ee
\figone
We note that there is no factor $L(\hat Y)$ in this divergence.
To obtain $B_{IJ}(\hat Y)$ in the second term of the divergence
(\ref{divergence}) we only need to consider the case
$\hat g_{\mu\nu} = \eta_{\mu\nu}$.
There are two kinds of divergent diagrams as shown in Fig.\ 1.
The internal line of the diagram (a) can be
\ba
{\rm (a1)} \A : \A \quad \VEV{hh}, \nonu
{\rm (a2)} \A : \A \quad \VEV{\xi\xi}.
\ea
Their divergences are
\ba
B_{IJ}^{\rm (a1)} \A = \A {1 \over 4\pi\epsilon} \left(
\hat G_{IJ} \hat L^{-1} + {5 \over 2} \partial_I \ln \hat L
\partial_J \ln \hat L \right), \nonu
B_{IJ}^{\rm (a2)} \A = \A -{1 \over 4\pi\epsilon} \left(
\hat R_{IJ} - \hat\Omega_{IAB} \hat\Omega_J{}^{AB} \right).
\label{bdiva}
\ea
The internal lines of the diagram (b) can be
\ba
{\rm (b1)} \A : \quad \A \VEV{hh},\ \VEV{hh}, \nonu
{\rm (b2)} \A : \quad \A \VEV{cb},\ \VEV{cb}, \nonu
{\rm (b3)} \A : \quad \A \VEV{hh},\ \VEV{\xi\xi}, \nonu
{\rm (b4)} \A : \quad \A \VEV{\xi\xi},\ \VEV{\xi\xi}.
\ea
Their divergences are
\ba
B_{IJ}^{\rm (b1)} \A = \A - {1 \over 2\pi\epsilon}
\partial_I \ln \hat L \partial_J \ln \hat L, \nonu
B_{IJ}^{\rm (b2)} \A = \A - {1 \over 8\pi\epsilon}
\partial_I \ln \hat L \partial_J \ln \hat L, \nonu
B_{IJ}^{\rm (b3)} \A = \A - {1 \over 4\pi\epsilon} \left(
\hat G_{IJ} \hat L^{-1} + \partial_I \ln \hat L
\partial_J \ln \hat L \right), \nonu
B_{IJ}^{\rm (b4)} \A = \A - {1 \over 8\pi\epsilon}
\hat\Omega_{IAB} \left(
\hat\Omega_J{}^{AB} - \hat\Omega_J{}^{BA} \right).
\label{bdivb}
\ea
The divergences of the $\Omega^2$ terms in (a2) and (b4)
cancel each other. A sum of the diagrams with $h_{\alpha\beta}$
propagators and the diagrams with ghost propagators gives non-zero
contributions to both of $A$ and $B_{IJ}$ in the divergence
(\ref{divergence}). From eqs.\ (\ref{adiv}), (\ref{bdiva}),
(\ref{bdivb}) we obtain the total one-loop divergence
\be
\Gamma_{\rm div}
= \int d^d x \sqrt{-\hat g} \biggl[
{24-N \over 24\pi\epsilon} \hat R^{(d)} - {1 \over 4\pi\epsilon}
\hat g^{\mu\nu} \partial_\mu \hat Y^I \partial_\nu \hat Y^J
\left( \hat R_{IJ} + \partial_I \ln \hat L
\partial_J \ln \hat L \right) \biggr].
\label{oneloopdiv}
\ee
The third term came from $h_{\mu\nu}$ and ghost diagrams.
It vanishes for the Einstein gravity, in which $L =$ constant.
\par
Using eqs.\ (\ref{idone}), (\ref{idtwo}), (\ref{idthree}) in the
appendix, the first two terms of eq.\ (\ref{oneloopdiv}) are shown
to be invariant under $\delta \hat Y^I$ in eq.\ (\ref{bweyl}).
Therefore, they are independent of the Liouville field $\hat\rho$
since $\delta \hat Y^I$ in eq.\ (\ref{bweyl}) is a shift of
the Liouville field. The third term changes under
$\delta \hat Y^I$ in eq.\ (\ref{bweyl}) as
\be
\delta_{\hat Y} \biggl[ \int d^d x \sqrt{-\hat g}
\hat g^{\mu\nu} \partial_\mu \hat Y^I \partial_\nu \hat Y^J
\partial_I \ln \hat L \partial_J \ln \hat L \biggr]
= 2 \epsilon \int d^d x \sqrt{-\hat g} \hat g^{\mu\nu}
\partial_\mu \sigma \partial_\nu \ln \hat L.
\ee
The $\hat\rho$-dependence is of order $\epsilon$. Therefore the
divergent part of (\ref{oneloopdiv}) is independent of the
Liouville field $\hat\rho$. Since the finite part of the counter terms
can be chosen at will, we shall choose a counter term which depends
on the background Liouville field $\hat\rho$ only through
$\hat g_{\mu\nu} \e^{-2\hat \rho}$. Then we should just replace
the background metric $\hat g_{\mu\nu}$ by the background physical
metric $\hat g_{\mu\nu} \e^{-2\hat \rho}$ which we shall denote
$g_{\mu\nu}$ without a hat $\, \hat{} \,$. This assures the general
coordinate invariance with respect to the physical metric
$g_{\mu\nu}=\hat g_{\mu\nu} \e^{-2\hat \rho}$. The invariance
under the background Weyl transformation (\ref{bweyl}) also
becomes manifest.
\par
Let us apply the result of one-loop divergence (\ref{oneloopdiv})
to our standard model (\ref{standardform}), which is rewritten
in terms of the Liouville field in eq.\ (\ref{dilatonaction}).
By computing the target space metric and curvature,
we obtain one-loop divergences of the dilaton gravity as
\be
\Gamma_{\rm div}
= \int d^d x \sqrt{- g} \biggl[
{24-N \over 24\pi\epsilon}  R^{(d)} - {1 \over 4\pi\epsilon}
 g^{\mu\nu} \partial_\mu \phi  \partial_\nu \phi
\left( N \left( \Phi'' + 2 \Phi' - (\Phi')^2 \right)
+ 4 \right) \biggr],
\label{diloneloopdiv}
\ee
where we have dropped to write a hat\ $\ \hat{}\ $\ for the background
field.
%
%
\newsection{Beta functions and fixed points}
The counter terms can be summarized with three types of
coefficients $A, B, C$
\be
S_{\rm counter} = - \mu^\epsilon \int d^d x \sqrt{-g} \left[
R^{(d)} A(\phi) + g^{\mu\nu} \partial_\mu \phi \partial_\nu \phi B(\phi)
- {1 \over 2} g^{\mu\nu} \partial_\mu X^i \partial_\nu X^i C(\phi)
\right].
\label{oneloopdivergence}
\ee
At one-loop level these coefficients are given by
\ba
A(\phi) \A = \A {24-N \over 24\pi\epsilon}, \qquad
C(\phi) =0, \nonu
B(\phi) \A = \A - {1 \over \pi\epsilon} + {N \over 4\pi\epsilon}
\left[ \left( \Phi'(\phi) \right)^2 - \Phi''(\phi)
- 2 \Phi'(\phi) \right].
\ea
The action including counter terms is expressed in terms of bare
quantities as
\ba
S_0 \A = \A S + S_{\rm counter} \nonu
\A = \A \int d^d x \sqrt{-g_0} \left[ {1 \over 16\pi G_0}
R^{(d)}_0 \e^{-2\phi_0} - {1 \over 2} g_0^{\mu\nu} \partial_\mu X_0^i
\partial_\nu X_0^i \e^{-2\Phi_0(\phi_0)} \right].
\ea
If we define
\ba
\Phi_0(\phi_0) \A = \A \Phi(\phi) + F(\phi) \qquad (F(0) = 0), \nonu
g_{0\mu\nu} \A = \A g_{\mu\nu} \e^{-2\Lambda(\phi)} \qquad
(\Lambda(0) = 0), \nonu
\phi_0 \A = \A \phi + f(\phi) \qquad (f(0) = 0), \nonu
X_0^i \A = \A \sqrt{Z} X^i,
\label{barequantities}
\ea
the action becomes
\ba
S_0 \A = \A \int d^d x \sqrt{-g} \biggl[
{1 \over 16\pi G_0} R^{(d)}
\e^{- 2\phi - \epsilon\Lambda - 2 f} \nonu
\A \A + {\epsilon+1 \over 16\pi G_0} \Bigl(
4 \Lambda' + \epsilon (\Lambda')^2 + 4 f' \Lambda' \Bigr)
\e^{- 2\phi - \epsilon\Lambda - 2 f}
g^{\mu\nu} \partial_\mu \phi \partial_\nu \phi \nonu
\A \A - {1 \over 2} Z \e^{- 2\Phi - \epsilon\Lambda - 2 F}
g^{\mu\nu} \partial_\mu X^i \partial_\nu X^i \biggr].
\ea
To reproduce the counter terms (\ref{oneloopdivergence}) we require
\ba
{1 \over 16\pi G_0} \e^{-2\phi - \epsilon\Lambda(\phi) - 2 f(\phi)}
\A = \A \mu^\epsilon \biggl[ {1 \over 16\pi G} \e^{-2\phi}
- A(\phi) \biggr], \nonu
{\epsilon+1 \over 16\pi G_0} \Bigl(
4 \Lambda' + \epsilon (\Lambda')^2 + 4 f' \Lambda' \Bigr)(\phi)
\e^{- 2\phi - \epsilon\Lambda(\phi) - 2 f(\phi)}
\A = \A - \mu^\epsilon B(\phi), \nonu
Z \e^{- 2\Phi(\phi) - \epsilon\Lambda(\phi) - 2 F(\phi)}
\A = \A \e^{- 2\Phi(\phi)} - C(\phi).
\ea
At one-loop level we can simplify these equations.
Using the fact that $A(\phi)$ is a constant, $C(\phi) = 0$ and
$\Lambda, f, F = O(G)$, and neglecting higher order terms in $G$,
we find that the above equations become
\ba
{1 \over 16\pi G_0} \e^{-2\phi - \epsilon\Lambda(\phi) - 2 f(\phi)}
\A = \A \mu^\epsilon \biggl( {1 \over 16\pi G} \e^{-2\phi}
- A \biggr), \nonu
{\epsilon+1 \over 4\pi G_0} \Lambda'(\phi) \e^{-2\phi}
\A = \A - \mu^\epsilon B(\phi), \nonu
\epsilon\Lambda(\phi) + 2 F(\phi) \A = \A 0, \qquad Z = 1.
\ea
Substituting the solution of these equations into
eq.\ (\ref{barequantities}) we obtain the relation between
the bare and renormalized quantities as
\ba
{1 \over G_0} \A = \A \mu^\epsilon \biggl( {1 \over G}
- 16\pi A \biggr), \nonu
\rho_0 \A = \A \rho - {4\pi G \over \epsilon+1}
\int_0^\phi d \phi' \e^{2\phi'} B(\phi'), \nonu
\phi_0 \A = \A \phi + 8\pi A G \left( \e^{2\phi} - 1 \right)
+ {2\pi \epsilon G \over \epsilon+1}
\int_0^\phi d \phi' \e^{2\phi'} B(\phi'), \nonu
\Phi_0(\phi_0) \A = \A \Phi(\phi)
+ {2\pi \epsilon G \over \epsilon+1}
\int_0^\phi d \phi' \e^{2\phi'} B(\phi'), \nonu
\A = \A \Phi(\phi_0)
- 8\pi A G \left( \e^{2\phi_0} - 1 \right) \Phi'(\phi_0) \nonu
\A \A - {2\pi \epsilon G \over \epsilon+1}
\left( \Phi'(\phi_0) - 1 \right)
\int_0^{\phi_0} d \phi' \e^{2\phi'} B(\phi').
\ea
\par
We find that the beta function $\beta$ and the anomalous dimension
$\gamma$ are functions of $\phi$ in general
\ba
\beta_G \A \equiv \A \mu {\partial G \over \partial \mu}
= \epsilon G - 16 \pi \epsilon A G^2,
\label{betafunction}
\nonu
\beta_\Phi(\phi_0) \A \equiv \A
\mu {\partial \Phi(\phi_0) \over \partial \mu} \nonu
\A = \A
8\pi \epsilon A G \left( \e^{2\phi_0} - 1 \right) \Phi'(\phi_0)
+ {2\pi \epsilon^2 G \over \epsilon+1}
\left( \Phi'(\phi_0) - 1 \right)
\int_0^{\phi_0} d \phi' \e^{2\phi'} B(\phi'), \nonu
\gamma_\rho \A \equiv \A \mu {\partial \rho \over \partial \mu}
= {4\pi \epsilon G \over \epsilon+1}
\int_0^\phi d \phi' \e^{2\phi'} B(\phi'), \nonu
\gamma_\phi \A \equiv \A \mu {\partial \phi \over \partial \mu}
= - 8\pi \epsilon A G \left( \e^{2\phi} - 1 \right)
- {2\pi \epsilon^2 G \over \epsilon+1}
\int_0^\phi d \phi' \e^{2\phi'} B(\phi').
\ea
The beta function $\beta_G$ for $G$ is similar to that of the
Einstein gravity. For $N < 24$, $G = 0$ is an infrared stable fixed
point and $G = G^*$ is an ultraviolet stable fixed point, where
\be
G^* = {3 \epsilon \over 2(24-N)}, \qquad
\beta_G (G^*)=0, \qquad \beta'_G (G^*) < 0.
\label{gravfp}
\ee
In order to find fixed points for the beta function $\beta_\Phi=0$,
we consider an ansatz
\be
\Phi(\phi) = \lambda \phi \qquad (\lambda = {\rm constant}).
\label{linearansatz}
\ee
Then the beta function becomes
\be
\beta_\Phi(\phi) = (\e^{2\phi}-1) \biggl[
{24-N \over 3}G\lambda + {\epsilon G \over \epsilon+1}
\left( {N \over 4}(\lambda^2-2\lambda)-1 \right) (\lambda-1)
\biggr].
\ee
The above beta function shows that fixed points can be obtained as
a solution of the cubic equation. We find that there is only one
fixed point for the real value of the parameter $\lambda$
\be
\Phi(\phi) = \lambda^* \phi, \qquad
\lambda^* = -{3\epsilon \over 24-N} + O(\epsilon^2).
\label{fp}
\ee
The other two solutions are purely imaginary and of order
$\epsilon^{-1/2}$
at the fixed point for the gravitational coupling constant $G^*$
\be
\lambda^* = \pm 2 i \sqrt{24-N \over 3N\epsilon} + O(\epsilon^0).
\label{imaginaryfp}
\ee
Since the imaginary solution corresponds to a wildly oscillating
dilaton coupling, we consider the real solution (\ref{fp}) to be
the only physical solution.
\par
Let us study stability of this fixed point (\ref{gravfp}),
(\ref{fp}). We expand
the beta functions near the fixed point
\be
G = G^* + \delta G, \qquad
\Phi = \lambda^* \phi + \delta\Phi,
\label{delphi}
\ee
assuming the fluctuations $\delta G$ and $\delta \Phi$ to be small.
By examining the beta functions near the fixed point
to first order in the fluctuation we find
\ba
\beta_G \A = \A - \epsilon \delta G, \nonu
\beta_\Phi \A = \A \epsilon (1-\e^{2\phi}) \delta G - {1 \over 2}
\epsilon (1-\e^{2\phi}) {d \over d\phi} \delta\Phi + O(\epsilon^2).
\label{deviation}
\ea
We can diagonalize these equation by considering the beta
function for
\be
\tilde\Phi(\phi) \equiv \Phi(\phi)
+ \left( 1 + {2\phi\e^{2\phi} \over 1-\e^{2\phi}} \right) G.
\ee
Then we obtain
\ba
\beta_G \A = \A - \epsilon \delta G, \nonu
\beta_{\tilde\Phi} \A = \A - {1 \over 2} \epsilon
(1-\e^{2\phi}) {d \over d\phi} \delta\tilde\Phi + O(\epsilon^2).
\label{diagdeviation}
\ea
Since $\e^{\phi}$ plays a role of the loop expansion parameter,
we shall only consider the region $-\infty < \phi \leq 0$,
in which $\e^{\phi} \leq 1$. If we define a variable $\psi$
\be
\psi = {1 \over 2} \ln (\e^{-2\phi} - 1), \qquad
\left\{
\begin{array}{l}
\psi \rightarrow + \infty \Longleftrightarrow
\phi \rightarrow - \infty, \\
\psi \rightarrow - \infty \Longleftrightarrow
\phi \rightarrow 0,
\end{array}
\right.
\label{defx}
\ee
the second equation in eq.\ (\ref{diagdeviation}) becomes
\be
\beta_{\tilde\Phi}
= {1 \over 2} \epsilon {d \over d\psi} \delta\tilde\Phi.
\ee
Eigenfunctions of the differential operator on the right hand
side is
\be
\delta \tilde \Phi =
\e^{\Lambda \psi} = (\e^{-2\phi} - 1)^{{1 \over 2}\Lambda}
\ee
with eigenvalues
\be
\beta_{\tilde\Phi}
= {1 \over 2} \epsilon \Lambda \delta\tilde\Phi.
\ee
The condition $\delta\Phi(\phi=0) = 0$ requires $\Lambda > 0$.
Therefore, the fixed point (\ref{gravfp}), (\ref{fp}) is not
ultraviolet stable in the direction $\delta\tilde\Phi$.
\par
We have also studied more general solutions without using the
linear ansatz (\ref{linearansatz}).
Since we are interested in the fixed point for the $\beta_\Phi$,
we take the gravitational coupling at the fixed point $G=G^*$.
We shall change variables from $\Phi$ to $\delta\Phi$ by
eq.\ (\ref{delphi}), but we no longer assume $\delta\Phi$ to be small.
We also use the variable $\psi$ defined in eq.\ (\ref{defx})
instead of $\phi$. By neglecting terms which are small for small
$\epsilon$, we obtain
\ba
\beta_{\delta \Phi}
\A = \A {24-N \over 3} G^* {d \delta \Phi \over d \psi}
+ {\epsilon G^* N \over 2(\epsilon+1)} \biggl[
1 + (1+\e^{-2\psi}) {d \delta\Phi \over d \psi} \biggr] \nonu
\A \A \times \int_{-\infty}^\psi d \psi' \biggl[
\e^{-2\psi'} \biggl( {d \delta\Phi \over d \psi'} \biggr)^2
- {d \over d \psi'} \biggl( \e^{-2\psi'}
{d \delta \Phi \over d \psi'} \biggr) \biggr].
\ea
We can explicitly solve the fixed point condition
$\beta_{\delta\Phi}=0$ and find that there are only three solutions
(\ref{fp}) and (\ref{imaginaryfp}) that has been obtained using
the linear ansatz (\ref{linearansatz}).
\par
The fixed point is found at (\ref{gravfp}) for the beta function
$\beta_G=0$ of the gravitational coupling and at (\ref{fp})
for the beta function $\beta_\Phi=0$ of the dilaton weight.
This fixed point is ultraviolet stable in the direction of $G$,
but is unstable in the direction of $\tilde\Phi$.
Therefore we can consider a renormalized theory with
the gravitational coupling constant near the fixed
point $G^*$, as long as we fine
tune the dilaton weight $\Phi$ to be precisely at the fixed point
$\delta \Phi=0$
\be
\Phi(\phi)=\lambda^* \phi
-\left(1+{2\phi{\rm e}^{2\phi} \over 1+{\rm e}^{2\phi}}\right)
(G-G^*).
\ee
\par
It is interesting to observe that the dilaton gravity theory at the
fixed point can be
recast into a form without matter coupling to the dilaton by
means of a nonsingular local Weyl rescaling, since the dilaton
weight $\Phi$ is of order $\epsilon$.
After the Weyl rescaling $g_{\mu\nu} \rightarrow g_{\mu\nu}\exp
\left(-{2\lambda^* \over \epsilon}\phi\right)$, we obtain an action
\ba
S \A = \A \int d^d x \sqrt{-g} \biggl[ {\mu^\epsilon \over 16\pi G^*}
\e^{-2(1-\lambda^*)\phi}
\left( R^{(d)} -{4(1+\epsilon) \over \epsilon}\lambda^*(2-\lambda^*)
g^{\mu\nu}\partial_{\mu}\phi\partial_{\nu}\phi \right) \nonu
\A \A - {1 \over 2} g^{\mu\nu} \partial_\mu X^i \partial_\nu X^j
\delta_{ij} \biggr],
\label{freexform}
\ea
This action is similar to the CGHS action (\ref{cghsaction})
since the coefficient in front of the dilaton kinetic term is
finite (${24 \over 24 -N}$) in the limit $\epsilon \rightarrow 0$
and is of the same sign as the CGHS action.
\par
%
%
%
\newsection{Discussion}
It has been observed in the Einstein gravity theory in
$2+\epsilon$ dimensions that one has to fix
the scale of the metric before one can meaningfully discuss
the renormalization of the gravitational coupling constant \cite{KN}.
One can fix the scale of the metric by introducing a reference
operator $\lambda \int d^d x \, O(x)$ with a coupling constant
$\lambda$ into the action.
The most convenient choice of the reference operator is a
gravitational dressing of an operator $\Psi$ which has a
dimensionless coupling constant $\lambda$ in the limit of
$\epsilon \rightarrow 0$, such as the Thirring interaction.
The gravitational dressing of such an operator takes the form of
\be
O=\e^{\alpha\kappa\rho}\Psi,
\ee
where the exponent $\alpha$ is of order $\epsilon$.
The merit of the decomposition (\ref{metricdec})
of the traceless mode $h_{\mu\nu}$ and the Liouville field $\rho$
is the fact that no one-loop divergence arises for gravitational
dressing of operators of the above form
\be
\VEV{\e^{\alpha\kappa\rho }}
 =  1 + {1 \over 2} (\kappa\alpha)^2 \VEV{\rho(0)\rho(0)}
+ \cdots ={\rm finite},
\ee
since the Liouville field propagator is of order $1/\epsilon$ and the
propagator at the coincidence point produces an additional
$1/\epsilon$ singularity both of which are canceled by
$\alpha^2=O(\epsilon^2)$.
Since the reference operator is automatically finite, one can
perform the renormalization of the gravitational coupling without
considering the reference operator explicitly.
\par
Now let us discuss renormalization of gravitationally dressed
operators in the dilaton gravity theory.
We need to introduce two reference operators in the action and use
them to fix the freedom of field redefinitions (zero modes of
$\Lambda(\phi)$ and $f(\phi)$) as we discussed
in sect.\ 2.
We would like to choose reference operators $O_i$ $(i=1,2)$ to be
gravitational dressing of operators $\Psi_i$ which have
dimensionless coupling constants $\lambda_i$ in the limit of
$\epsilon \rightarrow 0$, such as the Thirring interaction.
In the case of the dilaton gravity, however, the Liouville field
should be considered together with the dilaton because of the
mixing. Therefore we have a freedom in the gravitational dressing
to choose the amount of the dilaton dressing $\beta$ in addition
to the Liouville dressing $\alpha$.
\be
O_i=\e^{\alpha_i\kappa\rho + \beta_i\kappa\phi}\Psi_i.
\ee
If we take as reference operators those that become dimensionless in
the $\epsilon \rightarrow 0$ limit, we expect that the gravitational
dressing exponents $\alpha_i$ for the Liouville field to be
$O(\epsilon)$.
On the other hand, we have no particular reason to specify the
gravitational dressing exponent $\beta_i$ for the dilaton field.
Therefore $\beta_i$ is naturally expected to be $O(\epsilon^0)$.
These two reference operators serve to fix the scale of the metric
$g_{\mu\nu}$ (origin of $\rho$) and the origin of $\phi$.
\par
By using our propagators (\ref{propagator}),
we obtain divergences of the expectation value as
\ba
\VEV{\e^{\alpha\kappa\rho + \beta\kappa\phi}}
\A = \A 1 + {1 \over 2} \kappa^2 \VEV{(\alpha\rho + \beta\phi)^2}
+ \cdots \nonu
\A = \A 1 + {G \over \epsilon \mu^\epsilon}
(\alpha^2 + 2\alpha\beta) + O(\epsilon^0) + \cdots.
\ea
When the parameters satisfy $\alpha = O(\epsilon)$,
$\beta = O(\epsilon^0)$, there is no one-loop divergence.
Since these reference operators are automatically one-loop finite,
the gravitational coupling $G$ and the dilaton weight $\Phi(\phi)$
in our theory can be identified as those after the coupling
constants $\lambda_i$ for these reference operators are fixed.
\par
After writing this paper, we have received a preprint which has
worked out more details on the Einstein gravity
in $2+\epsilon$ dimensions \cite{AKKN}.
They showed that one can choose a renormalization group trajectory
which respects the general covariance, although their theory is
invariant only under volume preserving diffeomorphisms.
\par
\vspace{5mm}
%
%
Two of the authors (N.S. and S.K.) thank Yoshihisa Kitazawa for
interesting comments on our work and giving ref.\ \cite{AKKN}
prior to publication.
One of the authors (N.S.) also thanks Hikaru Kawai for a useful
discussion on the ($2+\epsilon$)-dimensional approach.
One of the authors (Y.T.) would like to thank the Theoretical
Physics Group of Imperial College for hospitality, and the Japan
Society for the Promotion of Science and the Royal Society for
a grant. This work is supported in part by Grant-in-Aid for
Scientific Research (S.K.) and (No.05640334) (N.S.), and
Grant-in-Aid for Scientific Research for Priority Areas
(No.05230019) (N.S.) from the Ministry of Education, Science
and Culture.
\vspace{5mm}
%
%
\def\numberbysectiona{\@addtoreset{equation}{section}
\def\theequation{A.\arabic{equation}}}
\numberbysectiona
\vspace{7mm}
\pagebreak[3]
\setcounter{section}{1}
\setcounter{equation}{0}
\setcounter{subsection}{0}
\setcounter{footnote}{0}
\begin{center}
{\large{\bf Appendix A. Conditions of background Weyl invariance}}
\end{center}
\nopagebreak
\medskip
\nopagebreak
\hspace{3mm}
In this appendix we discuss the conditions (\ref{bwcond}), which
have been obtained by requiring the background Weyl invariance
of the action (\ref{actiony}). First, let us derive several
identities which are used in the text. From the trace of the
first condition in eq.\ (\ref{bwcond}) we obtain
\be
D^I \partial_I L = {\rm constant}.
\label{idone}
\ee
By applying $D_K$ on the first condition in eq.\ (\ref{bwcond})
and antisymmetrizing the indices $K$ and $I$ we obtain
\be
\partial^M L R_{MJIK} = 0, \qquad \partial^J L R_{IJ} = 0.
\label{idtwo}
\ee
By applying $D^K$ on the first equation in eq.\ (\ref{idtwo})
and using the Bianchi identity and the first of eq.\ (\ref{bwcond})
we obtain
\be
\partial^K L D_K R_{IJ}  + D_I \partial^K L R_{KJ}
+ D_J \partial^K L R_{IK} = 0.
\label{idthree}
\ee
\par
Next, we shall obtain the general couplings $G_{IJ}$, $L$ which
satisfy the conditions (\ref{bwcond}). We split the target space
coordinates as $Y^I = (\rho, \phi, X^i)$ $(i = 1, \cdots, N)$.
By appropriately choosing the
coordinates  $Y^I$ we can put $G_{\rho I}$
to arbitrary functions in the target space.
This is most easily understood by noting that $G_{\rho\rho}$ and
$G_{\rho I}$ $(I\not= \rho)$ are analogous to the lapse
and the shift functions in the canonical formulation of general
relativity. We choose them as
\be
G_{I\rho} = 2 (\epsilon+1) \partial_I L.
\label{choice}
\ee
This choice corresponds to a coordinate system in which
the background Weyl transformation of $Y^I$ in eq.\ (\ref{bweyl})
becomes $\delta \rho = - \sigma$, $\delta Y^I = 0$ $(I\not= \rho)$.
Using eq.\ (\ref{choice}) the second condition in
eq.\ (\ref{bwcond}) becomes
\be
- \epsilon L - \partial_\rho L = 0.
\ee
The solution is
\be
L(Y) = \e^{-\epsilon\rho} \bar L(\phi, X),
\label{solone}
\ee
where $\bar L(\phi, X)$ is an arbitrary function of $\phi$ and $X^i$.
Substituting this into eq.\ (\ref{choice}) we obtain
\ba
G_{\rho\rho}(Y) \A = \A - 2 \epsilon(\epsilon+1) \e^{-\epsilon\rho}
\bar L(\phi, X), \nonu
G_{I\rho}(Y) \A = \A 2 (\epsilon+1) \e^{-\epsilon\rho}
\partial_I \bar L(\phi, X) \qquad (I\not=\rho).
\label{soltwo}
\ea
Then the first condition in eq.\ (\ref{bwcond}) becomes
\be
- \epsilon G_{IJ} - \partial_\rho G_{IJ} = 0.
\ee
This condition is automatically satisfied for $I=\rho$ or $J=\rho$
by eq.\ (\ref{soltwo}). The remaining cases
are solved by
\be
G_{IJ}(Y) = \e^{-\epsilon\rho} \bar G_{IJ}(\phi, X) \qquad
(I\not=\rho \ {\rm and} \ J\not=\rho),
\label{solthree}
\ee
where $\bar G_{ij}(\phi, X)$ is an arbitrary function of $\phi$ and
$X^i$.
Therefore, the general solution of eq.\ (\ref{bwcond}) is given by
eqs.\ (\ref{solone}), (\ref{soltwo}) and (\ref{solthree})
up to coordinate transformations in the target space. Substituting
this solution into eq.\ (\ref{actiony}) the action becomes
\ba
S \A =\A
{\mu^\epsilon \over 16\pi G}
\int d^d x \sqrt{-g} \biggl[
R^{(d)} \bar L(\phi, X)
- {1 \over 2} g^{\mu\nu} \partial_\mu \phi \partial_\nu \phi
\bar G_{\phi\phi}(\phi, X) \nonu
\A \A - g^{\mu\nu} \partial_\mu \phi \partial_\nu X^j
\bar G_{\phi j}(\phi, X)
- {1 \over 2} g^{\mu\nu} \partial_\mu X^i \partial_\nu X^j
\bar G_{ij}(\phi, X) \biggr],
\ea
where $g_{\mu\nu} = \tilde g_{\mu\nu} \e^{-2\rho}$.
This sigma model type action is the same as the general action
(\ref{generalaction}) after a trivial rescaling
of the target space metric $\bar G_{IJ}$.
Therefore the $\rho$-dependence is completely absorbed into
the physical metric $g_{\mu\nu}$.
%
\vspace{5mm}
%
\newcommand{\NP}[1]{{\it Nucl.\ Phys.\ }{\bf #1}}
\newcommand{\PL}[1]{{\it Phys.\ Lett.\ }{\bf #1}}
\newcommand{\CMP}[1]{{\it Commun.\ Math.\ Phys.\ }{\bf #1}}
\newcommand{\MPL}[1]{{\it Mod.\ Phys.\ Lett.\ }{\bf #1}}
\newcommand{\IJMP}[1]{{\it Int.\ J. Mod.\ Phys.\ }{\bf #1}}
\newcommand{\PR}[1]{{\it Phys.\ Rev.\ }{\bf #1}}
\newcommand{\PRL}[1]{{\it Phys.\ Rev.\ Lett.\ }{\bf #1}}
\newcommand{\PTP}[1]{{\it Prog.\ Theor.\ Phys.\ }{\bf #1}}
\newcommand{\PTPS}[1]{{\it Prog.\ Theor.\ Phys.\ Suppl.\ }{\bf #1}}
\newcommand{\AP}[1]{{\it Ann.\ Phys.\ }{\bf #1}}

\begin{thebibliography}{100}
%
\bibitem{WEI} S. Weinberg, in General Relativity, an Einstein
        Centenary Survey, eds.\ S.W. Hawking and W. Israel
        (Cambridge University Press, 1979) p.\ 790.
\bibitem{GK} R. Gastmans, R. Kallosh and C. Truffin, \NP{B133}
        (1978) 417.
\bibitem{CD} S.M. Christensen and M.J. Duff, \PL{B79} (1978) 213.
\bibitem{KN} H. Kawai and M. Ninomiya, \NP{B336} (1990) 115.
\bibitem{KKN} H. Kawai, Y. Kitazawa and M. Ninomiya,
        \NP{B393} (1993) 280.
\bibitem{KKNS} H. Kawai, Y. Kitazawa and M. Ninomiya,
        \NP{B404} (1993) 684.
\bibitem{KST} S. Kojima, N. Sakai and Y. Tanii, preprint TIT/HEP--238,
        Imperial/TP/93--94/8, hep--th/9311045, (1993), to appear in
        \IJMP{A}.
%
\bibitem{CGHS} C.G. Callan, S.B. Giddings, J. Harvey and
        A. Strominger, \PR{D45} (1992) R1005.
\bibitem{RUTS}
        J. Russo, L. Susskind and L. Thorlacius,
        \PL{B292} (1992) 13; \PR{D46} (1992) 3444;
        L. Susskind and L. Thorlacius, \NP{B382} (1992) 123;
        J. Russo and A.A. Tseytlin, \NP{B382} (1992) 259.
\bibitem{BDDO}
        T. Banks, A. Dabholkar, M. Douglas and M. O'Loughlin,
        \PR{D45} (1992) 3607;
        S. Hawking, \PRL{69} (1992) 406;
        B. Birnir, S.B. Giddings, J. Harvey and A. Strominger,
        \PR{D46} (1992) 638;
        E. Raiten, \PL{B289} (1992) 287;
        S.B. Giddings and W.M. Nelson, \PR{D46} (1992) 2486;
        S. Nojiri and I. Oda, \PL{B294} (1992) 317.
\bibitem{JACH} R. Jackiw, in Quantum theory of gravity,
        ed.\ S. Christensen (Hilger, Bristol, 1984) p. 403;
        C. Teitelboim, in  Quantum theory of gravity,
        ed.\ S. Christensen (Hilger, Bristol, 1984) p.327;
        K. Isler and C.A. Trugenberger, \PRL{63} (1989) 834;
        A.H. Chamseddine and D. Wyler, \PL{ B228} (1989) 75;
        \NP{ B340} (1990) 595;
        A.H. Chamseddine, \PL{B256} (1991) 379;
        \NP{ B368} (1992) 98;
        D. Cangemi and R. Jackiw, \PRL{69} (1992) 233.
%
\bibitem{DALW}
        S.P. De Alwis, \PL{B289} (1992) 278; \PL{B300} (1993) 330;
        A. Mikovic, \PL{B291} (1992) 19;
        S.B. Giddings and A. Strominger, \PR{D47} (1993) 2454;
        A. Strominger, \PR{D46} (1992) 4396;
        A. Bilal and C.G. Callan, \NP{ B394} (1993) 73;
        K. Hamada, \PL{B300} (1993) 322;
        K. Hamada and A. Tsuchiya, \IJMP{A8} (1993) 4897.
%
\bibitem{MASATAUC} Y. Matsumura, N. Sakai, Y. Tanii and T. Uchino,
        \MPL{A8} (1993) 1507.
%
\bibitem{TSE} E.S. Fradkin and A.A. Tseytlin, \NP{B261} (1985) 1;
        A.A. Tseytlin, \NP{B294} (1987) 383;
        \IJMP{A5} (1990) 1833.
%
\bibitem{BRZJLG} W.A. Bardeen, B.W. Lee and R.E. Shrock,
        \PR{D14} (1976) 985;
        E. Br\'ezin, J. Zinn-Justin and J.C. Guillou,
        \PR{D14} (1976) 2615.
\bibitem{DEWITT} B.S. De Witt, \PR{\bf 162} (1967) 1195, 1239;
        L.F. Abbott, \NP{B185} (1981) 189.
\bibitem{TV} G. 't Hooft and M. Veltman,
        {\it Ann.\ Inst.\ Henri Poincar\' e} {\bf 20} (1974) 69.
\bibitem{KU} T. Kugo and S. Uehara, \NP{B197} (1982) 378.
\bibitem{AFM} D. Friedan, \PRL{\bf 45} (1980) 1057;
        \AP{163} (1985) 318;
        L. Alvarez-Gaum\'e, D.Z. Freedman and S. Mukhi,
        \AP{\bf 134} (1981) 85; E. Braaten, T. Curtright
        and C. Zachos, \NP{B260} (1985) 630.
%
\bibitem{AKKN} T. Aida, Y. Kitazawa, H. Kawai and M. Ninomiya,
        preprint TIT/HEP--256, KEK--TH--395, YITP/U--94--13,
        hep-th/9404171 (1994).
%
\end{thebibliography}
\end{document}